\documentclass[%
 reprint, 
superscriptaddress, 
twocolumn,
nobibnotes,
 amsmath,amssymb,
 aps,
prb
]{revtex4-2}

\usepackage{siunitx}[=2021-04-09] 
\usepackage{spin-textures}
\usepackage{graphicx}
\usepackage{dcolumn}
\usepackage{tabularx}
\usepackage{physics}
\usepackage{float}
\usepackage{amsmath,amssymb,amsthm,bm}
\usepackage{subfigure}

\begin{document}

\preprint{APS/123-QED}

\title{Micromagnetics of magnetic chemical modulations in soft-magnetic cylindrical nanowires }

 \author{L.~Álvaro-Gómez}
 \email{laualv10@ucm.es}
 \affiliation{Univ. Grenoble Alpes, CNRS, CEA, Grenoble INP, SPINTEC, 38000 Grenoble, France.}
 \affiliation{Univ. Grenoble Alpes, CNRS, Institut Néel, 38000 Grenoble, France.}
 \affiliation{IMDEA Nanociencia, Campus de Cantoblanco, 28049 Madrid, Spain.}
 \affiliation{Dpto. de Física de Materiales, Universidad Complutense de Madrid, 28040 Madrid, Spain.}

\author{S.~Ruiz-Gómez}
\affiliation{Alba Synchrotron Light Facility, CELLS, 08290 Cerdanyola del Vallès, Barcelona, Spain.}
\affiliation{Max Planck Institute for Chemical Physics of Solids, 01187 Dresden, Germany.}

\author{C.~Fernández-González}
\affiliation{IMDEA Nanociencia, Campus de Cantoblanco, 28049 Madrid, Spain.}
\affiliation{Dpto. de Física de Materiales, Universidad Complutense de Madrid, 28040 Madrid, Spain.}

\author{M.~Schöbitz}%
\affiliation{Univ. Grenoble Alpes, CNRS, CEA, Grenoble INP, SPINTEC, 38000 Grenoble, France.}
\affiliation{Univ. Grenoble Alpes, CNRS, Institut Néel, 38000 Grenoble, France.}
\affiliation{Friedrich-Alexander Univ.\ Erlangen-N\"{u}rnberg, Inorganic Chemistry, Erlangen, Germany}

\author{N.~Mille}
\affiliation{Synchrotron SOLEIL, l’Orme des Merisiers, Saint-Aubin, FR-91192 Gif-sur-Yvette Cedex, France.}

\author{J.~Hurst}
\affiliation{Univ. Grenoble Alpes, CNRS, CEA, Grenoble INP, SPINTEC, 38000 Grenoble, France.}

\author{D.~Tiwari}
\affiliation{Univ. Grenoble Alpes, CNRS, CEA, Grenoble INP, SPINTEC, 38000 Grenoble, France.}

\author{A.~De Riz}
\affiliation{Univ. Grenoble Alpes, CNRS, CEA, Grenoble INP, SPINTEC, 38000 Grenoble, France.}

\author{I.M.~Andersen}
\affiliation{Centre d'Elaboration de Materiaux et d'Etudes Structurales, 31055, Toulouse, France.}

\author{J.~Bachmann}%
\affiliation{Friedrich-Alexander Univ.\ Erlangen-N\"{u}rnberg, Inorganic Chemistry, Erlangen, Germany}
\affiliation{Institute of Chemistry, Saint-Petersburg State Univ., St.\ Petersburg, Russia.}

\author{L.~Cagnon}
\affiliation{Univ. Grenoble Alpes, CNRS, Institut Néel, 38000 Grenoble, France.}

\author{M.~Foerster}
\affiliation{Alba Synchrotron Light Facility, CELLS, 08290 Cerdanyola del Vallès, Barcelona, Spain.}

\author{L.~Aballe}
\affiliation{Alba Synchrotron Light Facility, CELLS, 08290 Cerdanyola del Vallès, Barcelona, Spain.}

\author{R.~Belkhou}
\affiliation{Synchrotron SOLEIL, l’Orme des Merisiers, Saint-Aubin, FR-91192 Gif-sur-Yvette Cedex, France.}

\author{J.C~Toussaint}
\affiliation{Univ. Grenoble Alpes, CNRS, Institut Néel, 38000 Grenoble, France.}

\author{C.~Thirion}
\affiliation{Univ. Grenoble Alpes, CNRS, Institut Néel, 38000 Grenoble, France.}

\author{A.~Masseboeuf}
\affiliation{Univ. Grenoble Alpes, CNRS, CEA, Grenoble INP, SPINTEC, 38000 Grenoble, France.}

\author{D.~Gusakova}
\affiliation{Univ. Grenoble Alpes, CNRS, CEA, Grenoble INP, SPINTEC, 38000 Grenoble, France.}

\author{L.~Pérez}
\email{lucas.perez@ucm.es }
 \affiliation{IMDEA Nanociencia, Campus de Cantoblanco, 28049 Madrid, Spain.}
 \affiliation{Dpto. de Física de Materiales, Universidad Complutense de Madrid, 28040 Madrid, Spain.}
 \email{lucas.perez@ucm.es}

\author{O.~Fruchart}
\email{olivier.fruchart@cea.fr}
\affiliation{Univ. Grenoble Alpes, CNRS, CEA, Grenoble INP, SPINTEC, 38000 Grenoble, France.}
 \email{olivier.fruchart@cea.fr}

\date{\today}

\begin{abstract}
We analyze the micromagnetics of short longitudinal modulations of a high-magnetization material in cylindrical nanowires made of a soft-magnetic material of lower magnetization such as permalloy, combining magnetic microscopy, analytical modeling, and micromagnetic simulations. The mismatch of magnetization induces curling of magnetization around the axis in the modulations, in an attempt to screen the interfacial magnetic charges. The curling angle increases with modulation length, until a plateau is reached with nearly full charge screening for a specific length scale~$\Delta_\mathrm{mod}$, larger than the dipolar exchange length of any of the two materials. The curling circulation can be switched by the Oersted field arising from a charge current with typical magnitude $\SI{e12}{A/m^2}$ for a diameter of $\sim$\SI{100}{nm}, and reaching a maximum for $\Delta_\mathrm{mod}$.
\end{abstract}

\maketitle


\section{\label{sec:level1}Introduction}

Cylindrical magnetic nanowires can be synthesized rather easily by electrodeposition in insulating templates with pores, such as ion-track-etched polycarbonate membranes \cite{bib-FER1999a} and anodized aluminum foils and thin films \cite{bib-SOU2014}. Such wires have attracted intense and continuous interest over the past three decades, partially for being a  realization of a nearly one-dimensional magnetic system and thus a text-book case to understand fundamental phenomena. A wealth of phenomena have been investigated in cylindrical magnetic nanowires: current-perpendicular-to-plane giant magneto-resistance and measurements of spin diffusion length \cite{bib-PIR1996,bib-DUB1999,bib-EBE2000}, ferromagnetic resonance \cite{bib-EBE2001b}, competition between shape and magnetocrystalline anisotropy \cite{bib-DAR2005}, dipolar interactions in a dense and complex medium \cite{bib-DEL2009,bib-ROT2011,bib-FRU2011g}, nucleation at their apex \cite{bib-ZEN2000,bib-FOD2002,bib-ZEN2002,bib-WAN2008a}, statics and dynamics of domain-wall motion both under magnetic field and under a spin-polarized current  \cite{bib-EBE2000,bib-BIZ2013,bib-FRU2014,bib-FRU2016c,bib-FRU2019,bib-FRU2019b}.

The versatility of cylindrical nanowires is  high, with free choice of the composition and geometry, with diameters ranging from a few nanometers up to hundreds of nanometers, and lengths from a few tens of nanometers to tens of micrometers \cite{bib-FER1999a,bib-SOU2014,bib-VAZ2020}. In addition, it is possible to vary features along their length, either through the modulation of their diameter \cite{bib-PIT2011,bib-SAL2013,bib-SAL2018b,bib-BRA2016,Bran2021,bib-FER2018,ROD2016,bib-FRU2020,bib-NAS2019,bib-ALL2009,Dolocan2014,bib-FRU2018b} or modulation of their composition \cite{bib-IVA2016b,bib-BER2016,bib-BER2017,Ruiz2020,bib-VEG2012,Garcia2015,Rheem2007,Susano2016,bib-REY2016,bib-BOC2017, Wang2019}. The later is particularly interesting for engineering the longitudinal energy landscape and controlling domain-wall motion. From a fundamental point of view, this provides text-book cases of domain wall(DW) pinning on different types of defects \cite{bib-AHA1960,bib-AHA1962}, related to the long-standing problem called the Brown paradox \cite{bib-AHA1960,bib-AHA1962}. For applications it may provide ways to enhance coercivity, contributing to the search for rare-earth-free permanent-magnet materials \cite{GuzmanMinguez2020}, design non-reciprocal ratchet-type devices \cite{bib-BRA2018}, or provide a building block material for the concept of a domain-wall-based race-track memory \cite{bib-PAR2004,bib-PAR2008}.

For spintronics, modulations of composition are preferable to modulations of diameter as they avoid local changes in the current density \cite{bib-FRU2020}. There are already several theoretical and experimental investigations of nanowires modulated in composition, addressing the physics of magnetization reversal \cite{bib-MOH2016,bib-BRA2018,bib-IVA2017,bib-VEG2012,Chen2006,Corona2017, Kim2012}, domain-wall pinning \cite{bib-RUI2018,bib-MOH2016,bib-IVA2016b} or  field-driven domain wall motion  \cite{bib-RUI2018,bib-BRA2018,bib-BER2017,Ruiz2020}. Yet, the underlying physics of these modulations has not been examined in detail.

Here, we report the micromagnetics of chemical insertions within cylindrical nanowires. Although the insertions have higher magnetization than the rest of the nanowire, all materials considered are magnetically soft. Thus the materials of choice for the segments out of the chemical insertions are typically composed of Ni-rich, CoNi or FeNi alloys, which come with rather low magnetization. The nanowires are uniformly magnetized along their axis outside the modulations, that is, we are not considering domain walls.  We consider the practical case of Fe\(_{80}\)Ni\(_{20}\) chemical modulations periodically inserted in  Permalloy (Ni\(_{80}\)Fe\(_{20}\)), yet we draw conclusions in terms of reduced magnetic quantities when applicable, for the sake of generality. We combine simulations, micromagnetic modeling and experiments aiming to provide a comprehensive view, which we hope can be useful as a firm ground for the understanding of more complex situations involving chemical modulations, such as their interaction with magnetic domain walls.

\section{\label{sec:methods}Methods}
\subsection{\label{sec:expmethods}Experimental methods}

 Permalloy (Fe\(_{20}\)Ni\(_{80}\)) cylindrical nanowires with nanometer-sized modulations of composition Fe\(_{80}\)Ni\(_{20}\) were synthesized by template-assisted electrochemical deposition. Two different nanoporous anodic aluminum oxide (AAO) templates were prepared. First, AAO templates with pore diameter ranging between \SIrange{100}{140}{\nano\meter} were prepared  by one-step hard anodization of Al disks (Goodfellow, \SI{99.999}{\%} in purity) in a water-based solution of oxalic acid (\SI{0.3}{M}) and ethanol (\SI{0.9}{M}), applying \SI{140}{V} of anodization voltage at \SIrange{0}{1}{\celsius} for \SI{2.5}{h}. Second, AAO templates with pore diameter below 100~nm were prepared by two-step anodization in oxalic acid (\SI{0.3}{M}), applying \SI{60}{V} of anodization voltage at \SI{20}{\celsius} for \SI{15}{h} and \SI{3}{h} for the first and second anodization respectively. In both cases, the remaining Al was etched with an aqueous solution of CuCl$_2$ (\SI{0.74}{M}) and HCl (\SI{3.25}{M}), the oxide barrier was removed and the pores were opened to the final diameter with H$_3$PO$_4$ (5 {\% vol.}). In some cases we performed \SI{5}{nm} HfO$_2$ atomic layer deposition (ALD) coating of the pores to optimize the synthesis and prevent sample oxidation. However, this is not a key step because all the nanowires, also the ones without coating, have an oxidized CrO layer created by the solution used to release the nanowires from the alumina template. Therefore, at the end all the samples are protected from strong oxidation due to this

The nanowires were grown using pulse-plating electrodeposition under conditions similar to those described in Ref. \cite{bib-RUI2018}. The width of the pulses was varied to change the length $\ell$ of the chemical modulation from \SIrange{20}{150}{nm}, and the length of the Permalloy segments is a few hundreds of nanometers to several micrometers. Following growth, the templates were dissolved in H\(_{3}\)PO\(_{4}\)(\SI{0.4}{M}) and H\(_{2}\)CrO\(_{4}\)(\SI{0.2}{M}). Then, the nanowires were dispersed on Si substrates, either bulk or with a Si\(_{3}\)N\(_{4}\) window to allow for transmission microscopy. Nanowires were contacted electrically using laser lithography in order to allow current injection. Resistivity values obtained are as low as  \SI{20}{\micro\ohm\centi\metre} at room temperature in comparison with \SI{15}{\micro\ohm\centi\metre} of bulk Permalloy despite the material interfaces, allowing the injection of currents much higher than $\SI{e12}{\ampere\per\square\meter}$. The residual resistivity value measured at low temperature is \SI{10}{\micro\ohm\centi\metre}.

We used several types of magnetic microscopy techniques in order to access three-dimensional magnetization textures, all based on x-ray magnetic circular dichroism (XMCD): photo emission electron microscopy (PEEM) in shadow mode \cite{bib-FRU2014,bib-FRU2015c,bib-KIM2011b} and transmission x-Ray microscopy (TXM) \cite{Sorrentino2015} at CIRCE and MISTRAL beamlines of ALBA Synchrotron, Scanning transmission x-Ray microscopy (STXM)\cite{bib-BEL2015} and ptychography\cite{Pfeiffer2017,bib-DON2016,Mille2022} at HERMES beamline of SOLEIL synchrotron. Unless otherwise stated, we set the photon energy at the Fe L$_{3}$ absorption edge. The x-ray beam direction was typically nearly transverse to the nanowire axis, in order to be mostly sensitive to non-longitudinal components of the magnetization. For PEEM and TXM microscopy, series of about 64 images per polarity were taken and registered to increased the signal-to-noise ratio, while keeping under control the loss of spatial resolution due to drift during long acquisitions. For STXM, an image is reconstructed from a single scan. In the case of ptychography, an overlap of 30 \% of the illumination spots allowed a rather fast convergence of the iterative algorithm, and a reconstructed amplitude image of sub-\SI{15}{nm} resolution. Ptychographic reconstruction was carried out using the open-source PyNX software developed at the European Synchrotron Radiation Facility \cite{FavreNicolin2020}. Complementary measurements were performed by atomic and magnetic force microscopy (AFM and MFM respectively), as well as by scanning transmission electron microscopy (STEM) and holography.

\subsection{\label{sec:theormodel}Theoretical methods}
Micromagnetic simulations were performed using \textit{feeLLGood} \cite{bib-FEE,bib-STU2015,bib-ALO2014}, a homemade finite-element-based code that uses tetrahedrons for the discrete mesh. The system considered is a cylindrical nanowire made of two Permalloy segments of  \SI{100}{nm} (for a diameter $d=\SI{90}{\nano\meter}$)   or \SI{150}{nm} (for $d=\SI{130}{\nano\meter}$) length each, separated by a Fe\(_{80}\)Ni\(_{20}\) chemical modulation with length $\ell$ varied between 20 and \SI{120}{nm}. In order to mimic an infinite wire, the magnetic charges at the wire ends are removed numerically. The mesh size used was \SI{4}{nm}. The following material parameters were used: spontaneous magnetization $M_1 = \SI{8e5}{A/m}$ for the Permalloy segments, $M_2 = \SI{14e5}{{}A/m}$ for the Fe\(_{80}\)Ni\(_{20}\) modulation, $M_2 = \SI{12.5e5}{{}A/m}$ for the Fe\(_{65}\)Ni\(_{35}\) modulation, single exchange stiffness for all materials $A = \SI{1.3e-11}{J/m}$ and zero magnetocrystalline anisotropy. The reason to use a single value for exchange stiffness is to highlight the impact of the variation of $\Ms$ on the micromagnetics, which is the driving force for the phenomena reported here. The resulting dipolar exchange lengths $\DipolarExchangeLength = \sqrt{\frac{2A}{\muZero\Ms^2}}$ are $\SI{5.6}{nm}$, $\SI{3.3}{nm}$ and $\SI{3.7}{nm}$, respectively.   Finally, a damping coefficient equal to 1 was used for faster relaxation and dynamics. When considering the effect of an \OErsted field we consider a uniform and steady charge current flowing along the wire axis $z$, disregarding spin-transfer torque effects.

Post-processing scripts were used to extract quantitative information on the magnetization components from simulated and experimental data. From magnetic contrast images the degree of tilt of the magnetization was extracted based on the Beer-Lambert law as described in Ref. \cite{Schoebitz2021} (see Supplemental Material for a detailed description).

\begin{figure}[t]
\centering\includegraphics[width=\linewidth]{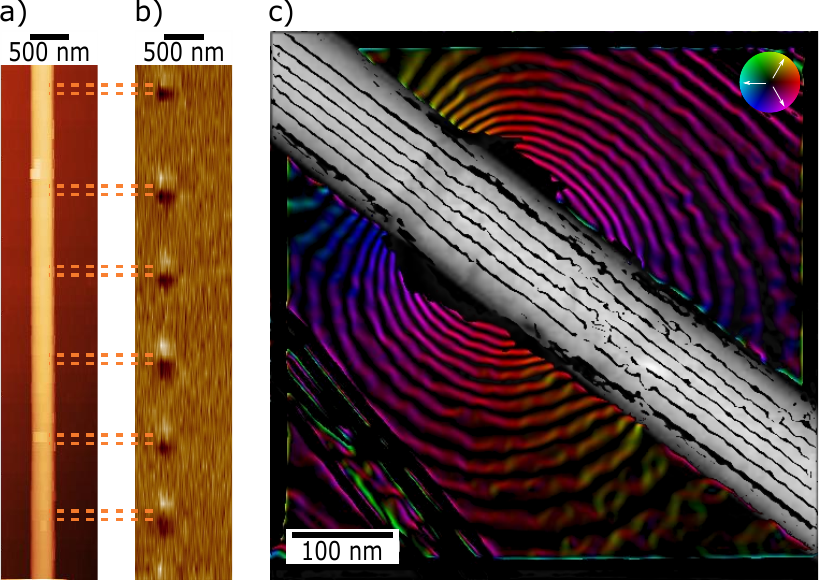}
\caption{\textbf{a)} Atomic Force Microscope (AFM) image of a chemically-modulated nanowire with $d=\SI{130}{\nano\meter}$ and $\ell=\SI{60}{nm}$. \textbf{b)} Magnetic Force Microscope (MFM) of a). Each modulation is highlighted by a dark/light bipolar contrast. Orange dashed lines indicate the location of the chemical modulations. \textbf{c)} Combination of STEM and holographic image of the nanowire in a). Black lines inside the wire show continuous induction field. The modulation is noticed by the black segment due to the difference in $Z$ contrast. Stray field lines arise from the modulation (directions sketched in the inset).  }
\label{fig:FIG1}
\end{figure}

\section{\label{sec:results}RESULTS}

\subsection{\label{sec:strayfield}Experimental evidence of stray field}

The insertion of chemical modulations with high magnetization within a Permalloy nanowire is expected to affect the magnetostatics of the system. Indeed, when the magnetization of the nanowire lies along its axis, as expected for soft magnetic materials in elongated structures, surface magnetic charges are expected to appear at the interface between the two materials, due to the mismatch of magnetization.

Fig. \ref{fig:FIG1}a-b show an AFM and a MFM image of a nanowire with $d=\SI{130}{nm}$ and $\ell=\SI{60}{nm}$, in a remanent state. In the MFM image, a bipolar magnetic contrast is observed at each modulation. This suggests the existence of opposite directions of the vertical magnetic stray field on either side of the modulation, which is consistent with the expectation of opposite magnetostatic charges at the two interfaces.

The stray field can be visualized directly with electron holography. The location of the modulations was determined first, by imaging in STEM, highlighting the $Z$ contrast. The stray field, imaged by electron holography, takes the form of a flux leakage at the modulation, opposite to the direction of magnetization in the wire~(Fig. \ref{fig:FIG1}c). For this specific case, the stray field is slightly shifted to the left due to a material defect on this particular wire. Note the continuity of the magnetic induction field $\vect{B}$ inside the wire, expected from Maxwell's equations. Yet, magnetic induction results both from magnetization and from the magnetostatic field, so that the details of the magnetization distribution cannot be easily obtained from this image. More is reported in the next section with other imaging techniques that are directly sensitive to the magnetization .

\begin{figure}[t]
\centering\includegraphics[width=\linewidth]{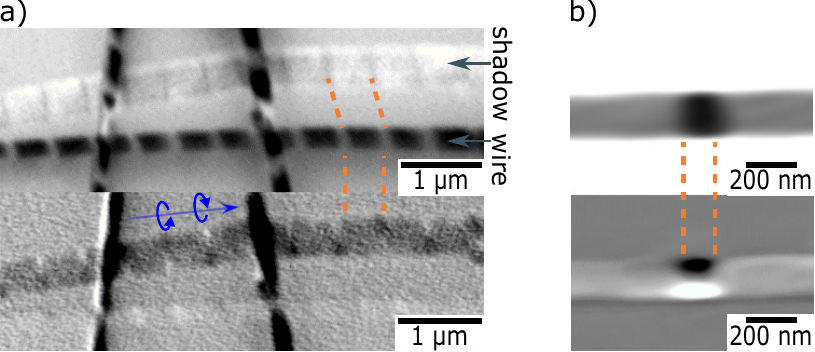}
\caption{X-Ray imaging Fe\(_{20}\)Ni\(_{80}\) cylindrical nanowires with Fe\(_{80}\)Ni\(_{20}\) chemical modulations.  \textbf{a)}  Top: chemical contrast image obtained by subtracting two shadow-PEEM XAS images taken at the Fe and Ni L$_{3}$ edges, for wire with $d=\SI{120}{nm}$ and $\ell=\SI{40}{nm}$ (black lines in the shadow).  The microscope focus was set at the shadow and the X-ray direction is almost perpendicular to the wire axis. The orange dashed lines guide the location of two chemical modulations. Bottom: XMCD corresponding to the top image, taken at the Fe L$_{3}$ edge. Blue arrows indicate magnetization direction. \textbf{b)}  Reconstructed amplitude ptychographic XAS (top) and  XMCD (bottom) of a wire with $d=\SI{160}{nm}$  and $\ell=\SI{150}{nm}$.
}
\label{fig:FIG2}
\end{figure}

\subsection{\label{sec:xmcd}Experimental evidence of curling}

Imaging techniques based on x-ray magnetic dichroism are complementary to electron holography. Indeed, the latter is sensitive to the two components  of magnetic induction perpendicular to the beam, while the former directly probes the magnetization component along the beam. Therefore, the combination of both provides a pretty good view of the three-dimensional distribution of magnetization.

Fig. \ref{fig:FIG2}a shows a shadow PEEM image of a nanowire with $d=\SI{120}{\nano\meter}$ $\ell=\SI{40}{nm}$. This specific wire is incidentally lying on top of two other wires, making its shadow fully visible. Shadows are stretched by about a factor of 3.5 along the beam direction due to the near grazing x-ray incidence (angle of \SI{16}{\degree}). The sample was rotated to have the nanowire almost perpendicular to the x-ray beam.

The top image is the subtraction of two x-ray absorption spectroscopy (XAS) images at the Fe and Ni L$_{3}$ edges, highlightling the changes in chemical composition. The Fe-rich chemical modulations appear bright~(more electrons emitted) and dark~(fewer photons transmitted) on the wire and the shadow, respectively. The bottom image is the corresponding XMCD image. The lack of magnetic contrast in the Permalloy segments reveals that the magnetization lies along the wire axis, consistent with the electron holography results~(Fig. \ref{fig:FIG1}c). However, magnetic contrast is observed at the modulations, mostly in the shadow as surface oxidation masks magnetic contrast on the wire surface. The bipolar dark and bright contrast at the modulations reflects components of magnetization parallel and anti-parallel to the X-ray beam on opposite sides of the wire. Such features are often called curling \cite{bib-FRE1957} in micromagnetism. The curved blue arrows in Fig. \ref{fig:FIG2}a shows that the sign of curling, which we will call circulation, can be either positive or negative in the same wire. The distribution of magnetization in the chemical modulations observed with higher spatial resolution by XMCD ptychography is shown in Fig. \ref{fig:FIG2}b. Note that in this case the aspect ratio along the transmitted x-ray beam is equal to one. We imaged nanowires with $d=$ \SIrange{70}{160}{nm} and $\ell=$ \SIrange{20}{150}{nm}, and evidenced curling in all of them. It should be noted that curling magnetization at the chemical modulations cannot be viewed as a 360$^{\circ}$ domain wall since the integration of change of magnetization angle along the axial direction is zero instead of 2$\pi$, which is the expected value for a 360$^{\circ}$  domain wall. 

\begin{figure}[t]
\centering\includegraphics[width=\linewidth]{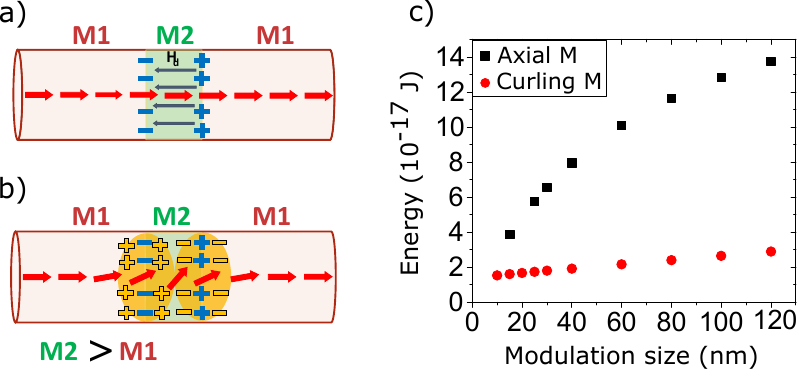}
\caption{Scheme of the magnetic charge distribution at a chemical modulation for \textbf{a)} axial  and \textbf{b)} curling  magnetization states, for $M_{2} > M_{1}$. The volume charges are shown in yellow, while the surface ones are shown with blue. \textbf{c)} shows the total magnetic energy of the axial and curling magnetization states, versus the modulation length for a diameter of 130 nm.}
\label{fig:FIG3}
\end{figure}

\subsection{\label{sec:analytics}Analytics to explain the physics of curling}
\label{sec-analytics}

Curling around the axis is a well-known feature of wires: for $d$ above typically seven times the dipolar exchange length, a mostly uniformly-magnetized wire ends up with such a curling feature at each end \cite{bib-ZEN2000,bib-WAN2008a,bib-FRU2018d}. Similar to the curling reversal mode from which the name was first coined \cite{bib-FRE1957}, this results from the driving force of surface charges, which appear at each end of the wire with areal density $\Ms$ in the case of uniform axial magnetization. For the following discussion, it is important to realize that curling does not remove magnetic charges from the system, whose integrated value is fixed by topology. Instead, curling converts part of the surface charges into volume charges $-\mathrm{div}\vectM$. These are on average further apart from each other than surface charges, thereby lowering the total magnetostatic energy of the system. In practice, the amount of curling is determined by the balance of gain in magnetostatic energy versus the cost in exchange energy. Another analogous situation is that of the vortex state in disks and short cylinders.
\begin{figure}[t]
\centering\includegraphics[width=0.9\linewidth]{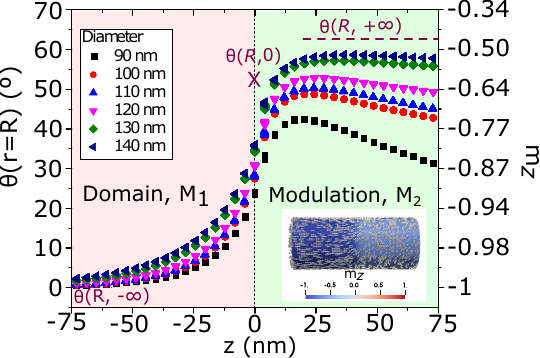}
\caption{Curling angle at the surface of the nanowire $\theta(R,z)$ obtained by micromagnetic simulations for various wire diameters. The prediction by the analytical model for curling at the interface and inside the modulation far from the interface are highlighted by a cross and a horizontal dashed line, respectively. The considered system is a Permalloy segment for $z<0$ and a $\mathrm{Fe}_{80}\mathrm{Ni}_{20}$  chemical modulation for $z>0$. Inset: magnetization state at the wire surface.
}
\label{fig:FIG4}
\end{figure}

A chemical modulation with high magnetization can be described in similar terms. The mismatch of spontaneous magnetization across the interfaces gives rise to surface charges of areal density $\sigma=\pm(M_{2}-M_{1})$ at the interfaces~(Fig. \ref{fig:FIG3}a). As a result, magnetization tends to curl at the modulation, reducing longitudinal magnetization, its interfacial mismatch, and the associated dipolar energy. The volume density of charge $\rho$ resulting from the gradient of longitudinal magnetization has a sign opposite to that of the nearby interfacial charges~(Fig. \ref{fig:FIG3}b), which can be understood as a screening effect. The initially dipolar nature of magnetostatics at the modulation tends to be replaced by an octopole, largely decreasing the extent of the resulting magnetostatic field, and thus its energy. It is important to remark that this occurs when the chemical modulation has a higher magnetization than the rest of the nanowire $(M_{2}>M_{1})$. For the opposite case, no screening mechanism may  occur since surface and volume charges have the same sign (for more details see the Supplementary Material). 

We propose several assumptions in order to model the physics of charge screening analytically in a simple manner and predict general features. In a first stage we consider just one interface, which would stand for a very long modulation: the domain material for $z<0$, and modulation material for $z>0$. Second, we consider a system with large diameter ($d\gg\DipolarExchangeLength$), which allows us to  neglect the radial and azimuthal contributions to exchange. The corolla of the second hypothesis is that the dominating contribution to the energy of the system is magnetostatics, so that as a third hypothesis we assume complete screening, \ie, exact compensation of the total amount of surface charge with volume charges of opposite sign. This is expected to bring the magnetostatic energy to a minimum. Fourth, we neglect any radial magnetization component.

\begin{figure}[t]
\centering\includegraphics[width=\linewidth]{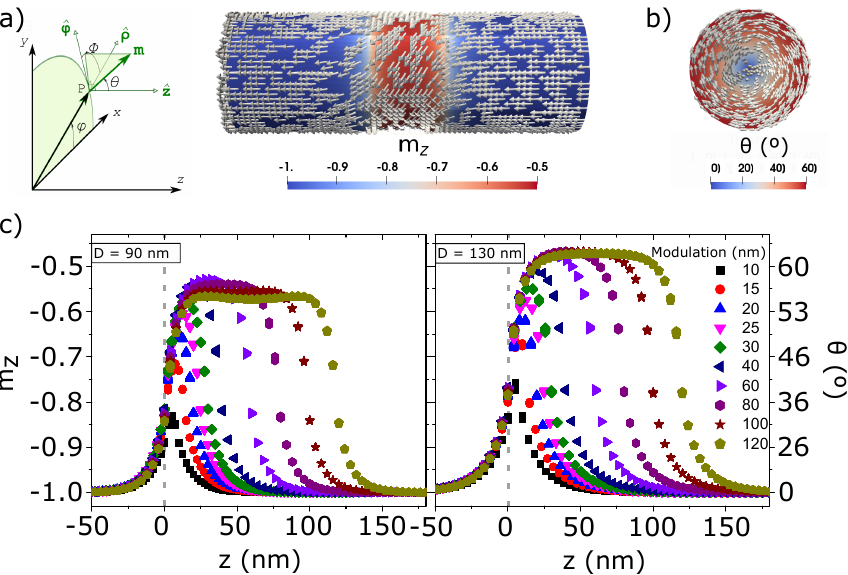}
\caption{Micromagnetic simulations of permalloy nanowires with a Fe\(_{80}\)Ni\(_{20}\) chemical modulation, at rest. \textbf{a)} Outer view and \textbf{b)} cross-section view at the center of the modulation, for $d=\SI{90}{nm}$ and $\ell=\SI{60}{nm}$. \textbf{c)}  Line profiles of unitless longitudinal magnetization $\vect{m}_{z}$ along the $\vect{z}$ direction at the surface of wire with a chemical modulation lengths  $\ell=$\SIrange{10}{120}{nm} for $d=\SI{90}{nm}$ (left) and $d=\SI{130}{nm}$ (right). The grey dashed line marks the start of the chemical modulation. }
\label{fig:FIG5}
\end{figure}

We define the curling angle $\theta$ as the angle of magnetization away from the axial direction and towards the azimuthal direction~(inset of Fig. \ref{fig:FIG5}a). As magnetization must remain longitudinal on the axis, the curling angle must be dependent on the distance to the axis. We use an ansatz to describe the variation of the curling angle $\theta$ along $r$:

\begin{equation}
\theta \left(r,z\right) = \theta(R,z) \sin{\left(\frac{\pi}{2}\frac{r}{R}\right)},
\label{eq:Ansatz}
\end{equation}
where $R$ is the wire radius, $r$ is the distance to the axis,  and $z$ is the coordinate along the wire axis. This ansatz accurately describes the spatial variation of $\theta$ along  $r$ \cite{bib-FRU2021,Schoebitz2021}, minimizing the (residual) exchange energy. The magnetization state of the system is then fully determined by the function $\theta(R,z)$, with $\theta(R,-\infty)=0$ in the axially magnetized domain. In this simplified example, $\theta(R,+\infty)$ stands for the final curling angle at the surface at a distance $z$ sufficiently far from the interface.

The Appendix provides the details of the model. In short, we use the following procedure: i)~We set $\theta(R,0)$ to a given value, not known at this stage. From this we compute the total surface charge at the interface, and the integrated volume charges on the left-hand side of the modulation. ii)~We assume an equal share of volume charges on either side of the interface, which allows us to derive the peripheral curling angle far inside the modulation, away from the interface, $\theta(R,+\infty)$, needed to establish these volume charges. This assumption is made with the aim of maximizing the gain provided by the screening of surface charges or, in other words, to obtain the largest screening for a given number of surface charges. iii)~We impose full screening of charges in a quadrupolar fashion, \ie, the integrated interfacial and volume charges compensate. This sets the values $\theta(R,0)$ and $\theta(R,+\infty)$ self-consistently. In the end we find:

\begin{eqnarray}
  \theta(R,0)&\approx&\sqrt{2.85\frac{M_{2}-M_{1}}{M_{2}+M_{1}}} \label{eq:angleInterface}\\
  \theta(R, +\infty) &\approx& \sqrt{2.85 \frac{M_{2}-M_{1}}{M_{2}}} \label{eq:angleInfty}
\end{eqnarray}
Based on the expansion $m_z(r=R,z)=\cos\theta(R,z)\approx1-\theta^2(R,z)/2$, we can rewrite these two equations for the longitudinal magnetization:
\begin{eqnarray}
  m_z(R,0)&\approx&1-1.42 \frac{M_{2}-M_{1}}{M_{2}+M_{1}} \label{eq:angleInterface2} \\
m_z(R, +\infty) &\approx& 1-1.42 \frac{M_{2}-M_{1}}{M_{2}} \label{eq:angleInfty2}
\end{eqnarray}
Note that the model yields the curling angle at the interface and deep inside the modulation~($z=+\infty$ in the model), however not the full function $\theta(R,z)$.  Also, the above expressions depend only on $\Ms$, because the model disregards exchange. Before assessing the validity of the model and hence its physical basis with micromagnetic simulations, let us make a few sanity checks. i)~There is no curling without a material modulation, \ie, for $M_2=M_1$ ii)~$M_1=0$ mimicks the situation of the apex of a cylindrical nanowire. In this situation we get $\theta(R,0)\approx1.69$, which is slightly more than $\pi/2\approx1.57$, the expectation to cancel surface charges. This comes from the hypothesis of the global compensation of charges: as no volume charge is created close to the axis for which magnetization remains longitudinal, more charges need to be created away from the axis. We will come back to this when comparing the model wnumerical micromagnetics.

Coming back to the specific case of Permalloy segments with modulations of inverted composition, on the basis of our model we expect the following: $\theta(R,0)\approx\SI{51}{\degree}$ [$m_z(R,0)\approx0.63$] and $\theta(R, +\infty)\approx\SI{61}{\degree}$ [$m_z(R,+\infty)\approx0.45$].

\subsection{\label{sec:micromagnetics}Micromagnetic simulation of curling}

The model reported in the previous paragraph intends to highlight the main physical ingredients driving curling, and to allow a quick estimation of the resulting angle. Here, we report a series of micromagnetic simulations to asses the validity of the model, refine its predictions quantitatively, and address the situation of a finite-length modulation, which cannot be tackled as easily through analytic modeling.

Fig. \ref{fig:FIG4} shows the curling profile at the surface $\theta(R,z)$, across a single interface between two materials with semi-infinite length derived from micromagnetic simulations. This is the situation considered in the model, whose predictions are indicated with horizontal dashed lines and a cross. It is clear that the model overestimates the curling angle, yet the agreement is better for larger diameter. This is because the model disregards exchange, while the energy cost coming from radial exchange decreases for larger diameter. We expect that other effects limiting curling are the fact that screening is non-local, as mentioned in the previous paragraph, balancing volume versus surface, and that magnetization cannot curl close to the wire axis. Yet, the very good quantitative agreement for curling inside the modulation for large diameter confirms that charge screening is indeed the key ingredient driving curling. The agreement for curling at the interface is not as good, although in both cases $\theta(R,0)>\theta(R,+\infty)/2$. The reason is that volume charges are related to the variation of longitudinal magnetization, which does not scale linearly with the curling angle, as the right $y$-axis in the plot reflects. Finally, note that curling tends to decrease in the modulation, when moving away from the interface. The reason is that the radial exchange cost of curling scales linearly with $\ell$ once the curling angle has reached a plateau, while it provides no further screening benefit. In the case of a semi-infinite medium, magnetization would progressively turn back fully axial for $z\rightarrow+\infty$ - an effect that is not reflected by the analytical model due to the disregard of exchange. The faster decay of curling for smaller diameter is consistent with this picture.

We now turn to the case of practical interest: a material modulation with finite length. Fig. \ref{fig:FIG5}a-b illustrate the surface and cross-sectional views of a simulated micromagnetic state at rest. The graphs in Fig. \ref{fig:FIG5}c-d show the peripheral $m_z(R,z)$ profile and curling angle, $\theta(R,z)$, versus axial position for two wire diameters, $\SI{90}{nm}$ and $\SI{130}{nm}$, each for a series of~$\ell$ values. The grey dashed line indicates the start of the chemical modulation at $z = 0$.

For long modulations the curling angle reaches a plateau, with a value very similar to the case of a single interface~(Fig. \ref{fig:FIG4}). Yet, the curling angle tends to decrease as $\ell$ increases due to an increasing cost in exchange. Thus, a maximum curling angle is found for a given modulation length: $\ell=\SI{60}{nm}$ for $d=\SI{90}{nm}$, with $|m_z| = 0.53$, , and $\ell=\SI{80}{nm}$ for $d=\SI{130}{nm}$, with $|m_z| = 0.47$. Note that $m_z(z)$ curves have nearly an inversion symmetry across each interface, unlike $\theta(R,z)$ curves. Since volume magnetic charges are related to the variation of $m_z$, this inversion symmetry shows that screening is largely symmetric.

The curling angle plateau is not reached for modulations shorter than typically \SIrange{30}{40}{nm}. There is no energy dependence of the chirality of the system, i.e., circulation versus longitudinal magnetization, so that both circulations have the same energy, which requires a certain distance to rotate magnetization.  This competition defines the dipolar exchange length~$\DipolarExchangeLength$, dictating a maximum gradient of magnetization direction. More precisely, with the definition $\DipolarExchangeLength = \sqrt{2A/(\mu_0 M_s^2)}$, magnetization may rotate by  $\SI{1}{\radian}$ over a distance~$\DipolarExchangeLength$. It can be seen on Fig. \ref{fig:FIG5}c-d that all $m_z(z)$ curves overlap at their rising stage in the Permalloy segment, even if the plateau is not reached. From the angular gradient in Fig. \ref{fig:FIG4}, we derive the characteristic length scale specific to a modulation as the distance that would be required to rotate magnetization by $\SI{1}{\radian}$, $\Delta_\mathrm{mod}\approx\SI{22}{nm}$, much larger than the dipolar exchange length of any of the two materials. We believe that the reason is twofold: first, the charges scale like $M_2-M_1$, not with the full magnetization; second, the screening effect on both sides of the interface further decreases the effective charge and thus the magnetostatic energy.

Micromagnetic simulations also allow us to derive the energy of the system for any distribution of magnetization. Fig. \ref{fig:FIG3}c shows the total micromagnetic energy versus modulation length, for longitudinal uniform magnetization and for the relaxed curling state for a 130 nm diameter wire. It is clear that curling allows to decrease energy, which however has a lower limit imposed by the residual magnetostatic energy and the resulting exchange energy.  The graph also illustrates how the gain is less for short modulations, consistent with the decrease in curling~(Fig. \ref{fig:FIG5}).

\subsection{\label{sec:axialField}Curling controlled by applied magnetic fields}

\subsubsection{\label{sec:expmethods}Response to an axial magnetic field}
\begin{figure}[t]
\centering\includegraphics[width=\linewidth]{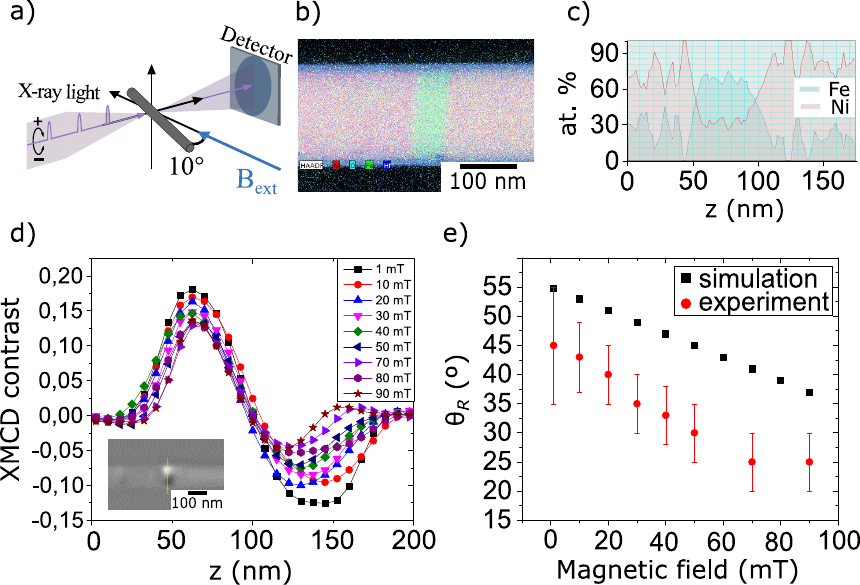}
\caption{Curling strength versus axial static magnetic field. \textbf{a)} STXM and Ptychography setup for a \SI{10}{\degree} rotation of the nanowire with respect the incident x-ray beam. \textbf{b)} and \textbf{c)} EELS map across the modulation. \textbf{d)}~XMCD contrast line profile across the modulation for several values of applied axial magnetic field. The inset shows the XMCD ptychography image and the line profile (yellow). The nanowire has $d=\SI{130}{nm}$ and $\ell=\SI{60}{nm}$. \textbf{e)}~Curling angle resulting from the analysis of the line profiles (red) and from micromagnetic simulations (black).}
\label{fig:FIG6}
\end{figure}

We performed high spatial resolution magnetic imaging by XMCD-ptychography in order to experimentally confirm the predictions of the previous section. Since the magnitude of curling depends on the materials parameters and the modulation length, which may all suffer from experimental errors, we studied both the curling angle at rest and its response to an external static magnetic field along the wire axis for a more robust comparison. We conducted experiments under a magnetic field  up to $\SI{90}{mT}$ applied perpendicular to the x-ray beam and thus almost along the  nanowire's axis. A small angular offset of  $\SI{10}{\degree}$ was chosen in order to have a small dichroic contribution from the axial magnetization and thus distinguish the magnetization direction in the domains (Fig. \ref{fig:FIG6}a).

The nanowire chosen for this study has $d=\SI{130}{nm}$ and $\ell=\SI{60}{nm}$. The experimental results and their analysis are summarized in Fig. \ref{fig:FIG6}. In Fig. \ref{fig:FIG6}d, we show the line profile of the XMCD contrast across the wire, centered on the modulation~(yellow line in the inset), for several applied magnetic fields. Its non-centro-symmetric shape results from sample rotation away from the plane perpendicular to the x-ray beam. The conversion of these curves into the curling angle is performed taking into account the intrinsic features~(exponential absorption along the x-ray path, 3D magnetization texture) and the experimental considerations~(finite size of the probe, background intensity) is explained in the Supplementary Material and in \cite{Schoebitz2021}. A crucial parameter of the model is the absorption coefficient of the material that depends on composition, which was independently obtained by electron-energy-loss- spectroscopy (EELS) measurements (Fig. \ref{fig:FIG6}b-c) as $\mathrm{Fe}_{65}\mathrm{Ni}_{35}$ at the modulation on this specific nanowire. In Fig. \ref{fig:FIG6}e, the evolution of  $\theta(R)$ as a function of the applied magnetic field is shown for data extracted from the experiments (red symbols) and from micromagnetic simulations (black symbols).
Curling decreases with applied field, however, it does not go to zero, owing to the large magnetostatic energy involved.

\subsubsection{\label{sec:OErsted}\OErsted field }

We have seen experimentally that curling in the modulations may be either clockwise or anticlockwise~(Fig. \ref{fig:FIG2}a). There is no energy dependence of the chirality of the system, i.e., circulation versus longitudinal magnetization, so that both circulations have the same energy. We expect that this degeneracy is lifted upon application of an electric current flowing along the wire, which gives rise to an \OErsted magnetic field that couples directly with curling, either parallel or antiparallel. This situation is analogous to the case of azimuthal curling around Bloch-point domain walls in nanowires, for which the \OErsted field has been proven to be crucial for the dynamic stability and switching \cite{bib-FRU2019b,bib-FRU2021}.

\begin{figure}[t]
\centering\includegraphics[width=0.9\linewidth]{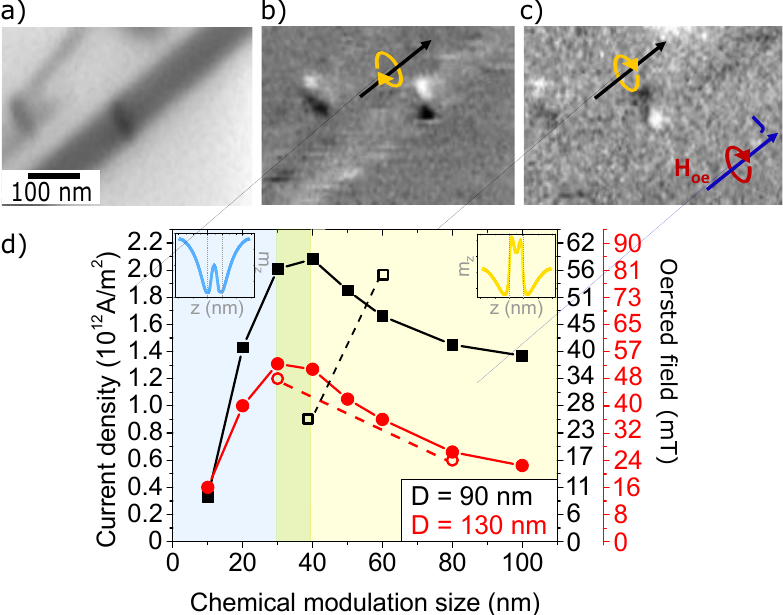}
\caption{Switching of the curling circulation in the modulation, driven by the \OErsted field. \textbf{a)} XAS Fe L$_{3}$ STXM image of a Permalloy nanowire electrically contacted, with $d=\SI{90}{nm}$ and $\ell=\SI{40}{nm}$. The nanowire at the top is not electrically ocntacted. \textbf{b)} XMCD image of \textbf{a)} in the initial state, revealing that curling in the modulation has negative circulation $C-$ (yellow arrow) with respect to the axial magnetization (black arrow). \textbf{c)} XMCD image after the application of a current pulse with density $j = \SI{-1.6e12}{A/m^2}$ and duration $\SI{5.6}{ns}$. The circulation switched to positive $C+$ circulation~(yellow arrow), which fits the circulation of the \OErsted field during the pulse of current~(red arrow). \textbf{d)} Threshold of current density to switch circulation, versus length of the chemical modulation. Simulation and experimental data is displayed with full and open simbols, for a system with $d=\SI{90}{nm}$ (black) and $d=\SI{130}{nm}$ (red) respectively. The y-axis at the right shows the corresponding \OErsted field at the external surface for $d=\SI{90}{nm}$ (values in black) and $d=\SI{130}{nm}$ (values in red). The green segment separates two regimes of switching. The insets in the first and second regime (blue and yellow background) show $m_z$ vs $z$ at the periphery of the wire before switching the circulation. }
\label{fig:FIG7}
\end{figure}

We conducted experiments on electrically-contacted nanowires, injecting current pulses of magnitude around $\SI{e12}{\ampere\per\square\meter}$, variable sign, and duration up to a few nanoseconds. The images were taken using XMCD-STXM or TXM, under static conditions before and after application of every current pulse. We evidence the existence of a threshold for the current density, above which the sign of curling in the modulation switches deterministically, if not initially parallel to the \OErsted field. The switching is illustrated in Fig. \ref{fig:FIG7} for a wire with $d=\SI{90}{nm}$ and $\ell=\SI{40}{nm}$. Fig. \ref{fig:FIG7}a shows an absorption image, highlighting the position and width of the modulation. The initial state, revealed by XMCD in b, has a negative circulation $C-$ (yellow arrow) with respect the magnetization direction (black arrow). After the application of a current pulse with magnitude $j=\SI{-1.6e12}{A/m^2}$ and duration $\SI{5.6}{ns}$, circulation has switched to a positive circulation state $C+$ (Fig. \ref{fig:FIG7}c, yellow arrow). This is consistent with the sign of the \OErsted field associated with the charge current~(red arrow). We evidenced no dependence of the switching with pulse duration for lengths \SIrange{1}{20}{ns}. Open black and red symbols in Fig. \ref{fig:FIG7}d show the experimental dependence of the threshold current versus modulation length~$\ell$ for $d=\SI{90}{nm}$ and $d=\SI{130}{nm}$ respectively. The calculation of the current density that flows through the wire was done taken into account the attenuation and losses of the electric pulse (see Supplemental Material).

We performed micromagnetic simulations to shed light on the switching mechanism. We considered only the effect of the \OErsted field, disregarding spin-transfer torque~(which was previously shown to be negligible \cite{bib-FRU2021}), as well as Joule heating, which plays a key role in magnetization dynamics for high current density values or long pulses. In our experiment we estimated a temperature rise of $\SI{100}{\celsius}$ for a current density  of magnitude $j = \SI{1e12}{A/m^2}$, pulse duration $\SI{5}{ns}$ and 90 nm diameter.  The initial magnetization state prepared has a negative circulation with respect to the axial magnetization direction $m_z$, as in Fig. \ref{fig:FIG5}a, antiparallel to the \OErsted field that will be applied. We need a criterion to determine the minimum current density $j_\mathrm{c}$ required to switch the circulation. Indeed, since any computation time is finite, a long incubation time may be misinterpreted as absence of circulation switching. A standard method in micromagnetism to address this issue is to rely on the scaling of a parameter, such as the switching time $t_\mathrm{sw}$ above the threshold, versus the driving parameter, here the current density~$j$. The switching time was defined by the first occurrence of $\theta = 0$ at the surface of the wire, which reflects the transition towards positive circulation. We expect and indeed evidence that $1/t_\mathrm{sw}$ scales nearly linearly with $|j-j_\mathrm{c}|$\cite{bib-FRU2021}, from which $j_\mathrm{c}$ is determined.

We considered lengths $\ell$ ranging from 20 to $\SI{100}{\nano\meter}$ and two diameters: $d=\SI{90}{nm}$ and $d=\SI{130}{nm}$. The results are summarized in Fig. \ref{fig:FIG7}d, along with the experimental results. One can see that the switching current is higher for smaller diameter. While the micromagnetic energetics may play a role, the dominant effect is probably the magnitude of \OErsted field. Indeed, it rises linearly with radius for a given density of current, thus promoting switching. In addition, the variation with modulation length is non-monotonous, displaying a maximum of switching current around $\ell=\SI{40}{\nano\meter}$ for $d=\SI{90}{nm}$ and around $\ell=\SI{30}{\nano\meter}$ for $d=\SI{130}{nm}$. This hints at the existence of two distinct switching regimes, which we understand as follows. For short modulations, the curling angle (Fig. \ref{fig:FIG5}) as well as the energy barrier (Fig. \ref{fig:FIG3}c) decrease, so that switching is easier. For long modulations, the curling angle presents a minimum at the center of the modulation, which we understand as follows: While curling at the interfaces is driven by significant magnetostatic energy and the tendency for charge screening, curling at the center is imposed only through exchange with the two interfacial regions, which decays with modulation length once the plateau is reached. Thus, the central region is expected to be a weak locus for nucleation. Examination of snapshots of the magnetization distribution during the switching process, confirms that it occurs largely as a coherent process in cylindrical coordinates for short modulations, in terms of $\theta(r,t)$, while for long modulations the switching process is not coherent but rather with a profile $\theta(r,z,t)$ that indicates nucleation at the center of the modulation. Insets in Fig. \ref{fig:FIG7}d show $m_z$ profiles at the periphery of the wire before the circulation reversal. In the short modulation regime (blue blackground) $m_z$ decreases more rapidily at the edges of the modulation than at the center (grey dashed lines in the inset) whereas in the long modulation regime (yellow background) this occurs at the center of the modulation. The high value of $m_z$ out of the modulation reflects the tilt of the Permalloy domains due to the \OErsted field (for more details see the Supplementary Material). We pointed out in the micromagnetics of the modulation at rest, that the characteristic length scale here is $\Delta_\mathrm{mod}\approx\SI{22}{nm}$, in the case of $\mathrm{Fe}_{80}\mathrm{Ni}_{20}$ modulations within $\mathrm{Fe}_{20}\mathrm{Ni}_{80}$ segments. This means, magnetization may rotate by $\SI{180}{\degree}$ over a distance~$\approx\pi\Delta_\mathrm{mod}$. As the plateau for curling reaches $\SI{60}{\degree}=\SI{\pi/3}{\radian}$, so that the total rotation is twice this value, we expect that the length required is $(2*\pi/3)\Delta_\mathrm{mod}\approx\SI{46}{nm}$. This is consistent with the modulation width giving rise to the maximum of switching current~(Fig. \ref{fig:FIG7}d), which confirms the relevance of a specific length scale for modulations, $\Delta_\mathrm{mod}$.

The order of magnitude for the switching current is consistent between simulations and experiments. However, while the experimental trend for $d=\SI{130}{nm}$ is consistent with the above, for $d=\SI{90}{nm}$ it is found to be opposite. Reasons can be several-fold. First, in the simulations we did not consider a distinct value for exchange stiffness in the modulation material, while a larger exchange is expected in practice for $\mathrm{Fe}_{80}\mathrm{Ni}_{20}$, which shall increase $\Delta_\mathrm{mod}$. Second, we have neglected thermal activation~(expected to have more impact for shorter modulations and smaller diameters) as well as heating during the current pulse. Third, the modulation length and materials composition suffer from experimental uncertainties.

\subsection{\label{sec:conclusion}Conclusion}

Combining experiments, analytical modeling and micromagnetic simulations, we provide a rather consistent qualitative and quantitative view of the micromagnetics of high-magnetization longitudinal material modulations in cylindrical nanowires. Their key feature is the spontaneous occurrence of curling around the axis, driven by the need to create volume magnetic charges and screen the interfacial charges between different materials. This replaces a net charge by a quadrupolar distribution for an interface, and the dipole of the total modulation by an octopole, thereby sharply decreasing the magnetostatic energy of the system. An analogy can be drawn with vortex states in disks and short cylinders, sharing the same feature of magnetization distribution, however with different quantitative energetics. We evidence a specific length scale, $\Delta_\mathrm{mod}$, a few times larger than the dipolar exchange length of any of the two materials and expected to scale with $\sqrt{2A/\muZero(M_2-M_1)^2}$. For modulations shorter than typically $2\Delta_\mathrm{mod}$ the charge screening is partial, and switching of the curling sign with an \OErsted field occurs in a coherent fashion. For modulations longer than this, the charge screening is nearly complete, the curling reaches an angle that can be predicted analytically, and switching of the curling sign with an \OErsted field occurs incoherently, nucleating from the center of the modulation and expanding towards its interfaces. The switching current reaches its maximum at the cross-over between the two regimes. We believe that this knowledge of the micromagnetics at modulations is important in order to understand more complex situations, such as their interaction with magnetic domain walls.

\subsection{\label{sec:conclusion}Acknowledgments}

We acknowledge support from the French RENATECH network, the team of the Nanofab platform (CNRS Néel institut), and the CNRS-CEA METSA French network (FR CNRS 3507). The project received financial support from the French National Research Agency (Grant No. ANR-17-CE24-0017-01) (MATEMAC-3D), from the Spanish MCIN/AEI/ 10.13039/501100011033 under grant PID2020-117024GB-C43 as well as by the Comunidad de Madrid (Spain) through
the project NanoMagCOST (CM S2018/NMT-4321).
Sandra Ruiz-G\'omez gratefully acknowledges the financial suppport of the Alouexander von Humboldt foundation.

\subsection{\label{sec:conclusion}Data availability}
Associated data is also available on \href{https://doi.org/10.5281/zenodo.6532308}{https://doi.org/10.5281/zenodo.6532308}.
\section{\label{sec:annex}Annex}

Here, we detail intermediate steps supporting the analytical model, whose main features are described in section~\ref{sec-analytics}.

Throughout the derivation of the analytical model, we make use of the following integrations:
\begin{equation}
\int r\sin^2\left({\frac\pi2\frac rR}\right)\diff r=\frac{r^2}{4}-\frac{Rr}{2\pi}\sin\left({\frac{\pi r}{R}}\right)-\frac{R^2}{2\pi^2}\cos\left({\frac{\pi r}{R}}\right).
\label{eq:annex-integral-rhosin2}
\end{equation}
\begin{equation}
\begin{aligned}
\int r\cos\theta(r,z)\diff r=& \frac{r^2}{2}- \frac{\theta^2(R,z)}{2}\times \\
& \left[{\frac{r^2}{4}-\frac{Rr}{2\pi}\sin\left({\frac{\pi r}{R}}\right)-\frac{R^2}{2\pi^2}\cos\left({\frac{\pi r}{R}}\right)}\right].
\label{eq:annex-integral-rcostheta}
\end{aligned}
\end{equation}
\begin{equation}
\left<{m_z(r,z)}\right>_r\approx 1 - \theta^2(R,z)\left({\frac14 + \frac{1}{\pi^2}}\right)
\label{eq:annex-mz-averaged}
\end{equation}
\begin{equation}
\left<{m_z^2(r,z)}\right>_r\approx 1 - \frac{\theta^2(R,z)}{2}\left({1 + \frac{4}{\pi^2}}\right)
\label{eq:annex-mz2-averaged}
\end{equation}
The latter two equations relate to the radial areal average, and make use of the expansion of $\cos\theta$ to second order, \ie, up to $\theta^2$.

The distribution of magnetic charges at the interface between the two materials reads:
\begin{equation}
\sigma(r)=(M_2-M_1)\cos\theta(r,0).
\label{eq:annex-charge-at-interface}
\end{equation}
Upon integration on the full disk cross-section, we find the resulting total charge, expanding the cosine up to second order:
\begin{equation}
\Sigma=-\pi R^2 (M_2-M_1)\left[{1 - \theta^2(R,z)\left({\frac14 + \frac{1}{\pi^2}}\right)}\right]
\label{eq:annex-total-charge-at-interface}
\end{equation}
The distribution of volume magnetic charges reads:
\begin{equation}
\rho=-\fracpartial{M_z(z)}{z}=-M_{1,2}\fracpartial{\left[{\cos\theta(r,z)}\right]}{z}
\label{eq:annex-volume-charges}
\end{equation}
Upon integration along both the radius~$r$ and the longitudinal coordinate~$z$, we derive the total volume charge:
\begin{equation}
\begin{aligned}
P=& \pi R^2\left({\frac14 + \frac{1}{\pi^2}}\right) \times \\
  & \left\{{ M_1\theta^2(R,0)+M_2 \left[{\theta^2(R,+\infty)-\theta^2(R,0)}\right] }\right\}.
\label{eq:annex-integral-rcostheta}
\end{aligned}
\end{equation}
In the above, the first term in brackets on the right-hand side stands for volume charges in material~1, while the second term stands for charges in material~2. We now make the assumption that volume charges are equally shared on either sides of the interface, to achieve a quadrupolar distribution and minimize the energy of the system. This sets the condition:
\begin{equation}
\theta^2(R,0)=\frac{M_2}{M_1+M_2}\;\theta^2(R,+\infty) ,
\label{eq:annex-link-angles-infty-interface}
\end{equation}
and allows to simplify the expression for the total volume charges:
\begin{equation}
\begin{aligned}
P&=\pi R^2 \left({\frac14 + \frac{1}{\pi^2}}\right) \times \frac{2 M_1 M_2}{M_1+M_2}\;\theta^2(R,+\infty) \\
  & = \pi R^2 \left({\frac14 + \frac{1}{\pi^2}}\right) \times 2 M_1 \theta^2(R,0) \label{eq:annex-link-angles-infty-interface},
\end{aligned}
\end{equation}
We now equal the total interface and volume charges in absolute value~[\eqnref{eq:annex-total-charge-at-interface} and \eqnref{eq:annex-link-angles-infty-interface}], assuming full screening. This leads to the following expressions:
\begin{eqnarray}
  \theta^2(R,+\infty) & = & \frac{M_2-M_1}{M_2}\;\frac{1}{\frac14 + \frac{1}{\pi^2}}
  \label{eq:annex-angle-infty-result} \\
  \theta^2(R,0) & = & \frac{M_2-M_1}{M_2+M_1}\;\frac{1}{\frac14 + \frac{1}{\pi^2}}
  \label{eq:annex-angle-infty-result} \\
\end{eqnarray}
\eqnref{eq:angleInterface} and \eqnref{eq:angleInfty} directly derive from the above, with a numerical approximation for $1/(1/4 + 1/\pi^2)$.

\bibliography{Fruche8,Lauracomp} 

\end{document}


\preprint{APS/123-QED}
\title{Supplementary material: Micromagnetics of magnetic chemical modulations in soft-magnetic cylindrical nanowires }

  \author{L.~Álvaro-Gómez}
 \email{laualv10@ucm.es}
 \affiliation{Univ. Grenoble Alpes, CNRS, CEA, Grenoble INP, SPINTEC, 38000 Grenoble, France.}
 \affiliation{Univ. Grenoble Alpes, CNRS, Institut Néel, 38000 Grenoble, France.}
 \affiliation{IMDEA Nanociencia, Campus de Cantoblanco, 28049 Madrid, Spain.}
 \affiliation{Dpto. de Física de Materiales, Universidad Complutense de Madrid, 28040 Madrid, Spain.}

\author{S.~Ruiz-Gómez}
\affiliation{Alba Synchrotron Light Facility, CELLS, 08290 Cerdanyola del Vallès, Barcelona, Spain.}
\affiliation{Max Planck Institute for Chemical Physics of Solids, 01187 Dresden, Germany.}

\author{C.~Fernández-González}
\affiliation{IMDEA Nanociencia, Campus de Cantoblanco, 28049 Madrid, Spain.}
\affiliation{Dpto. de Física de Materiales, Universidad Complutense de Madrid, 28040 Madrid, Spain.}

\author{M.~Schöbitz}%
\affiliation{Univ. Grenoble Alpes, CNRS, CEA, Grenoble INP, SPINTEC, 38000 Grenoble, France.}
\affiliation{Univ. Grenoble Alpes, CNRS, Institut Néel, 38000 Grenoble, France.}
\affiliation{Friedrich-Alexander Univ.\ Erlangen-N\"{u}rnberg, Inorganic Chemistry, Erlangen, Germany}

\author{N.~Mille}
\affiliation{Synchrotron SOLEIL, l’Orme des Merisiers, Saint-Aubin, FR-91192 Gif-sur-Yvette Cedex, France.}

\author{J.~Hurst}
\affiliation{Univ. Grenoble Alpes, CNRS, CEA, Grenoble INP, SPINTEC, 38000 Grenoble, France.}

\author{D.~Tiwari}
\affiliation{Univ. Grenoble Alpes, CNRS, CEA, Grenoble INP, SPINTEC, 38000 Grenoble, France.}

\author{A.~De Riz}
\affiliation{Univ. Grenoble Alpes, CNRS, CEA, Grenoble INP, SPINTEC, 38000 Grenoble, France.}

\author{I.M.~Andersen}
\affiliation{Centre d'Elaboration de Materiaux et d'Etudes Structurales, 31055, Toulouse, France.}

\author{J.~Bachmann}%
\affiliation{Friedrich-Alexander Univ.\ Erlangen-N\"{u}rnberg, Inorganic Chemistry, Erlangen, Germany}
\affiliation{Institute of Chemistry, Saint-Petersburg State Univ., St.\ Petersburg, Russia.}

\author{L.~Cagnon}
\affiliation{Univ. Grenoble Alpes, CNRS, Institut Néel, 38000 Grenoble, France.}

\author{M.~Foerster}
\affiliation{Alba Synchrotron Light Facility, CELLS, 08290 Cerdanyola del Vallès, Barcelona, Spain.}

\author{L.~Aballe}
\affiliation{Alba Synchrotron Light Facility, CELLS, 08290 Cerdanyola del Vallès, Barcelona, Spain.}

\author{R.~Belkhou}
\affiliation{Synchrotron SOLEIL, l’Orme des Merisiers, Saint-Aubin, FR-91192 Gif-sur-Yvette Cedex, France.}

\author{J.C~Toussaint}
\affiliation{Univ. Grenoble Alpes, CNRS, Institut Néel, 38000 Grenoble, France.}

\author{C.~Thirion}
\affiliation{Univ. Grenoble Alpes, CNRS, Institut Néel, 38000 Grenoble, France.}

\author{A.~Masseboeuf}
\affiliation{Univ. Grenoble Alpes, CNRS, CEA, Grenoble INP, SPINTEC, 38000 Grenoble, France.}

\author{D.~Gusakova}
\affiliation{Univ. Grenoble Alpes, CNRS, CEA, Grenoble INP, SPINTEC, 38000 Grenoble, France.}

\author{L.~Pérez}
\email{lucas.perez@ucm.es }
 \affiliation{IMDEA Nanociencia, Campus de Cantoblanco, 28049 Madrid, Spain.}
 \affiliation{Dpto. de Física de Materiales, Universidad Complutense de Madrid, 28040 Madrid, Spain.}
 \email{lucas.perez@ucm.es}

\author{O.~Fruchart}
\email{olivier.fruchart@cea.fr}
\affiliation{Univ. Grenoble Alpes, CNRS, CEA, Grenoble INP, SPINTEC, 38000 Grenoble, France.}
 \email{olivier.fruchart@cea.fr}

\date{\today}

\maketitle

\section*{X\lowercase{-ray absorption spectroscopy: correlation between experiments and theory }}

\begin{figure}[h]
\centering\includegraphics{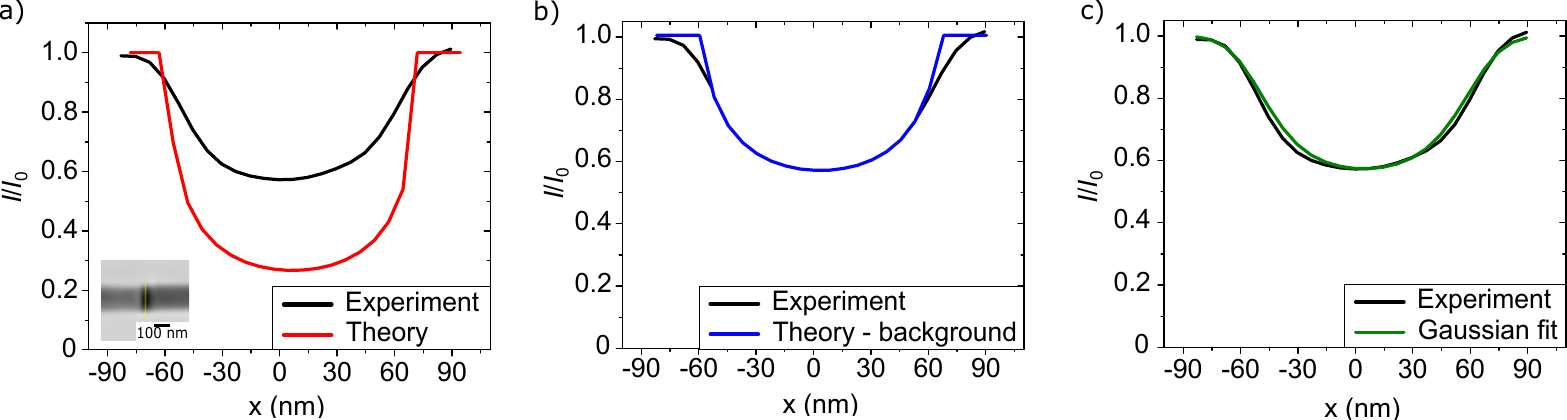}
\caption{Analysis of XAS and XMCD reconstructed ptychographic image to extract the spot size and curling angle. a) XAS normalized intensity profile across the cross-section of the modulation for a nanowire with $d=\SI{130}{nm}$ and $\ell=\SI{60}{nm}$ obtained experimentally (black curve). Yellow line in the inset shows the location of the line profile. The theoretical transmitted intensity is plotted by the red curve. b) Theoretical transmitted intensity after subtracting the background signal (blue) vs experimental data (black). c) Theoretical data in b) convoluted with a Gaussian whose width ($\sigma$ = 12 nm) results from the fit with the experimental data.}
\label{fig:FIGS1}
\end{figure}

There is not direct correlation between the value of magnetic contrast and direction of magnetization since the absorption of x-rays follows an exponential behavior based on Beer-Lambert law:
\begin{equation}
I\left(x\right) = I_0 \exp \left[-\mu t\left(x\right)\right]
\label{eq:XAS-intensity}
\end{equation}
where $I\left(x\right)$ is the intensity of the transmitted x-rays through a material,  $t\left(x\right)$ the thickness of the material and $\mu$ the absorptivity coefficient. Moreover, the experiments have a finite resolution or spot size that needs to be considered. It can be modeled by a Gaussian:
\begin{equation}
I_\mathrm{spot}\left(x\right) = \exp \left(-  x^2 / 2 \sigma^2  \right)  
\label{eq:Spot}
\end{equation}
where $\sigma$ is the standard deviation of the Gaussian, which is a measurement of the spot size.   To extract quantitative information of the experimental data we have developed a post-processing code. 
Several approaches were followed to determine the absorptivity coefficient at the modulation $\mu_{\text{Fe}_{80}\text{Ni}_{20}}$ related to this particular experiment. The main issue to face is the fact that magnetization is not uniform across the modulation and thus the component of the magnetization along the x-ray beam ($\hat{\vect{k}}\cdot\vect{m})$ is unknown. To solve this, we decided to consider the theoretical value $\mu_{th,\text{Fe}_{80}\text{Ni}_{20}}$ and extract the background signal associated to experimental limitations. For this, we calculate the background signal, $I_{b, \text{XAS}}$, by subtracting the minimum value of the experimental profile of the transmitted intensity from the minimum calculated theoretically (black and red curve in Fig. \ref{fig:FIGS1}a. Then we add $I_{b, \text{XAS}}$ value to the theoretical intensity (Fig. \ref{fig:FIGS1}b. To extract the spot size, a convolution of the theoretical profile with a Gaussian of unknown width was done followed by a fit with the experimental data (Fig. \ref{fig:FIGS1}c). In this specific case, $\sigma$ = 12 nm was obtained.

The calculation of the curling angle $\theta(R)$ from the magnetic contrast of the images was done using the post processing code describe in \cite{Schoebitz2021} where the input parameters were the wire diameter (130~nm), $\mu$, $\sigma$, the sample rotation ($\alpha = \beta = \SI{10}{\degree}$) and the background signal. In this case the background signal was determined by $I_{b,\text{XMCD}} =  I_{b,\text{XAS}} \times I_{\text{XMCD,out}}$ where $I_\text{XMCD,out}$ is the XMCD intensity out of the wire.  The theoretical absorptivity coefficients used were $\mu_{\text{Fe}_{65}\text{Ni}_{35}}$ = 0.045 nm$^{-1}$ and $\Delta \mu_{\text{Fe}_{65}\text{Ni}_{35}}$ = 0.026 nm$^{-1}$. In this case, a curling angle of  $\theta(R)$ = \SI{32}{\degree} for 50 mT of axial magnetic field was extracted. Furthermore, the $r^2$ for this fit is $0.94$.

\section*{C\lowercase{alculation of the current density switching threshold}}

\begin{figure}[h]
\centering\includegraphics{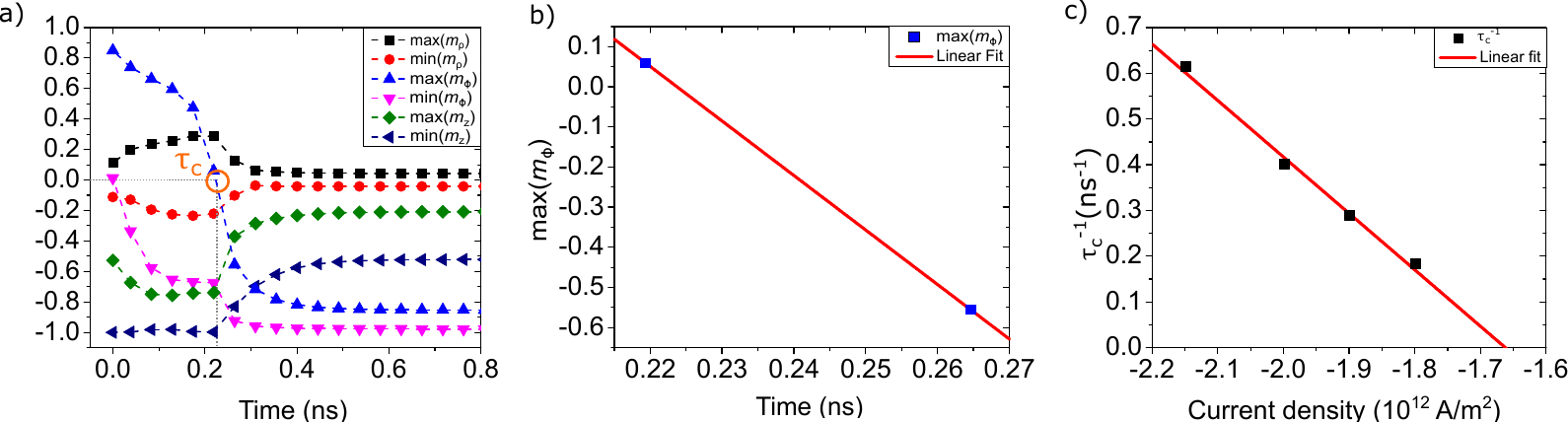}
\caption{Method to extract threshold current density of switching. a) Maximum and minimum value of $m_{z}$, $m_{\rho}$ and $m_{\phi}$ plotted over time of a line profile at the surface of the wire before and after the switching of the circulation. The nanowire diameter is $d=\SI{90}{nm}$  and $\ell=\SI{60}{nm}$ and the current density $j = \SI{3e12}{A/m^2}$. The switching time $\tau_{c}$ is determined by max($m_{\phi}$) = 0. b) Linear interpolation between the values before and after max($m_{\phi}$) = 0 to extract exact value of $\tau_{c}$. c) Linear interpolation of the values of $\tau_{c}^{-1}$ as a function of $j_{c}$. Interception with the horizontal axis gives the threshold current density.  }
\label{fig:FIGS2}
\end{figure}

A method based on the scaling of a parameter was used to determine the minimum current density to switch the circulation at the modulation. We have used the critical time $\tau_{c}$ to switch above the threshold as parameter . Before this, a criterion to define $\tau_{c}$ needs to be set. In Fig. \ref{fig:FIGS2}a, the time evolution of the the maximum and minimum values of a line profile at the surface of the wire of $m_{z}$, $m_{\rho}$ and $m_{\phi}$ are plotted. In this specific case, the system was a nanowire with $d=\SI{90}{nm}$  and $\ell=\SI{60}{nm}$ and $j = \SI{3e12}{A/m^2}$. We decided to determine $\tau_{c}$ as the time when the azimuthal component changes its sign which is at max($m_{\phi}$) = 0 (orange circle in Fig. \ref{fig:FIGS2}a. The extraction of $\tau_{c}$ was done by doing a linear fit between the values before and after max($m_{\phi}$) = 0 (Fig. \ref{fig:FIGS2}b). This procedure was applied for several of $j_{c}$  to obtain several  values of $\tau_{c}$. The threshold current density corresponds to the j value for $\tau_{c}^{-1}$=0 and can be obtained by plotting $\tau_{c}^{-1}$ as a function of $j_{c}$ (see Fig. \ref{fig:FIGS2}c).

\section*{I\lowercase{dentification of two switching regimes}}

\begin{figure}
\centering\includegraphics[]{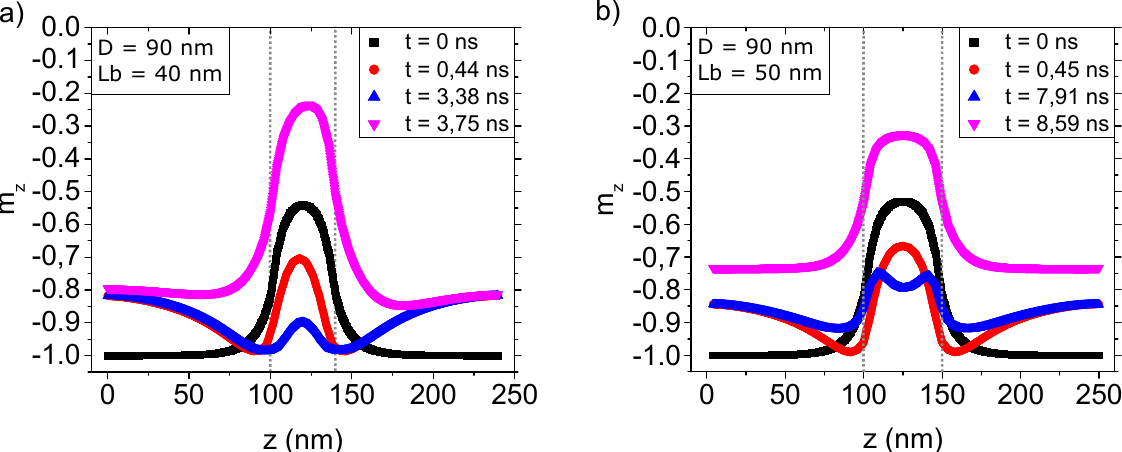}
\caption{Axial magnetization evolution during the switching mechanism. a) First regime: $m_{z}$ vs $z$ profiles over time at the surface of a nanowire of $d=\SI{90}{nm}$ and $\ell=\SI{40}{nm}$ under the application of a current pulse $j = \SI{-2.10e12}{A/m^2}$. The grey dashed lines set the location of the modulation. The profile in pink shows $m_{z}$ after the reversal to a circulation parallel to the \OErsted field. b) Second regime: same as in a) for   nanowire of $d=\SI{90}{nm}$ and $\ell=\SI{50}{nm}$ under the application of a current pulse $j = \SI{-1.9e12}{A/m^2}$. }
\label{fig:FIGS3}
\end{figure}

The switching of circulation driven by the \OErsted field associated to a current pulse can occur at two different regimes, in the  first regime the reversal is a coherent process throughout the modulation and  in the second regime it is not coherent since the switching starts at the center of the modulation. The identification of the value of $\ell$ limiting both regimes was set by the maximum threshold current density, $j_{c,th}$, needed for reversal. This occurred at $\ell=\SI{40}{nm}$ for $d=\SI{90}{nm}$ wire and at $\ell=\SI{30}{nm}$ for $d=\SI{130}{nm}$ wire. However, since $j_{c,th}$ values for these modulation lengths are quite similar we decided to look in more detail the switching mechanism. For this, we analyzed over time the $m_{z}$ profiles along the axial direction at the periphery of the wire. For the case of the first regime (short modulation lengths), the peak of the profile decreases homogeneously. This is shown in Fig. \ref{fig:FIGS3}a for a nanowire of $d=\SI{90}{nm}$ and $\ell=\SI{40}{nm}$. The black curve corresponds to the $m_{z}$ profile at rest, then the red and blue curve are the profiles while injecting a pulse of $j = \SI{-2.10e12}{A/m^2}$ and the pink curve refers to the profile under the same pulse but after the reversal of the circulation. The tilt of the Permalloy domains by the \OErsted field can be observed. Also the location of the modulation is set by the grey dashed lines. The same situation is explored in Fig. \ref{fig:FIGS3}b for a nanowire of $d=\SI{90}{nm}$ and $\ell=\SI{50}{nm}$ under the application of a pulse of amplitude $j = \SI{-1.9e12}{A/m^2}$. In this case, the $m_{z}$ profile before switching (blue curve) reaches its minimum at the center of the modulation. We observed this shape of the profile for $\ell\geq\SI{50}{nm}$, which sets a criteria to correlate a switching mechanism within the second regime.

\section*{C\lowercase{urrent density calculation from oscilloscope }}

Calculating the  the current density that flows trough the wire is a crucial step in the analysis of the experiments. To do so, the transmission, $t$, and attenuation, $\beta$, coefficients need to be calculated from the output voltage from the pulse generator $V_{output}$, the reflected voltage from the sample $V_{reflected}$ and the transmitted voltage trough the sample $V_{transmitted}$. The voltage, both output and reflected/transmitted, measured trough 50 Ohm (either transmission cable or end load), are directly proportional to the current. These coefficients are described as:

\begin{equation}
t = \frac{V_{transmitted}}{V_{transmitted} + V_{reflected} }
\label{}
\end{equation}

\begin{equation}
\beta = \frac{V_{transmitted} + V_{reflected} }{V_{output}}
\label{}
\end{equation}

\begin{figure}[H]
\centering\includegraphics[]{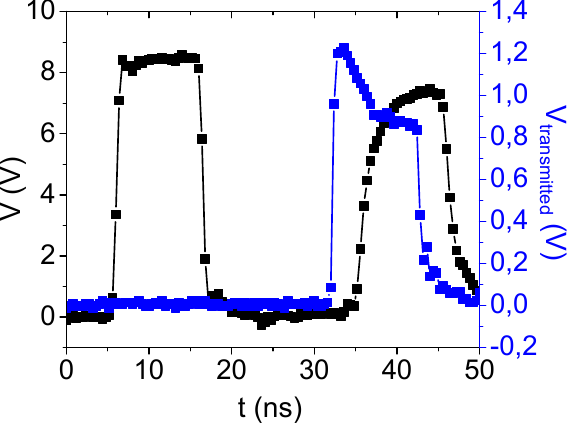}
\caption{Characterization of a 10 ns current pulse trough a nanowire of $d=\SI{135}{nm}$, $length=\SI{17}{um}$, resistance $R=\SI{282}{\ohm}$ and resistivity $\rho=\SI{23}{\micro\ohm\centi\metre}$. The black curve shows an output voltage of $V_{output} = \SI{8.5}{\volt}$  and reflected voltage $V_{reflected} = \SI{6.25}{\volt}$. The blue curve shows a transmitted voltage of $V_{transmitted} = \SI{1.25}{\volt}$.  }
\label{fig:FIGS4}
\end{figure}

As an example, Fig. \ref{fig:FIGS4} shows the voltage evolution over time for a 10 ns pulse show $V_{output} = \SI{8.5}{\volt}$, $V_{reflected} = \SI{6.25}{\volt}$ and $V_{transmitted} = \SI{1.25}{\volt}$, which results in $t = 0.17$ and $\beta = 0.88$. The decrease and increase over time in the transmitted and reflected signal is due to the increase of temperature due to Joule heating that changes the electrical resistance of the wire. We consider that the dynamics of our events happens within the first nanoseconds and we always consider the maximum transmitted signal accordingly. This nanowire had the following properties $d=\SI{135}{nm}$, $length=\SI{17}{um}$, resistance $R=\SI{282}{\ohm}$ and resistivity $\rho=\SI{23}{\micro\ohm\centi\metre}$. The associated current density is $j = V_{transmitted} / (\sqrt{\beta} \times A \times R_{osciloscope}) =  \SI{1.25}{\volt} / (\SI{0.94} \times \SI{14314}{nm^{2}} \times \SI{50}{\ohm} ) = \SI{1.86e12}{A/m^2}$. It is important to take into account that the value of $\beta$ strongly depends on the  setup used for each experiment ($\beta = 0.30$ for the PEEM setup used in this work) and it is crucial to consider the appropriate value for the calculation of the real current density. 

\section*{C\lowercase{ase of lower magnetization at the chemical modulation }}
\begin{figure}[H]
\centering\includegraphics[scale=1.4]{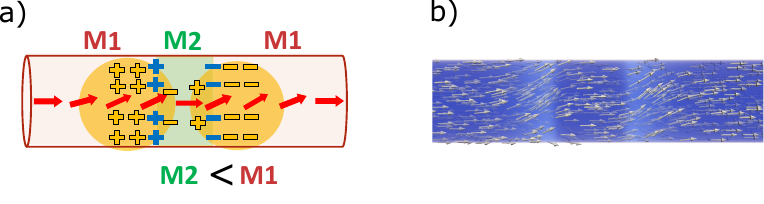}
\caption{a) Scheme of magnetic charges distribution for a chemical modulation of lower magnetization $(M_{2}<M_{1})$. Surface and volume charges are represented in blue and yellow respectively. Red arrows represent magnetization direction. b) Micromagnetic simulations of a 90 nm diameter ${\text{Fe}_{80}\text{Ni}_{20}}$ nanowire with a 100 nm long ${\text{Fe}_{20}\text{Ni}_{80}}$ chemical modulation. White arrows represent magnetization direction and colorcode axial magnetization.}
\label{fig:FIGS5}
\end{figure}

In this manuscript we have not considered the case of a chemical insertion of lower magnetization since, for this situation the physics is significantly different. However, we would like to address in this section the magnetic charges distribution for the case where $M_{2}<M_{1}$. In this situation, a nanowire with positive axial magnetization as represented by the red arrows in Fig. \ref{fig:FIGS5} a) have positive surface charges ($\sigma=\pm(M_{2}-M_{1})$ and positive volume charge distribution  ($\rho=-\fracpartial{M_z(z)}{z}$) at the first interface. At the second interface both magnetic charges have negative sign. The equality in sign of the magnetic charges implies that no screening mechanism can occur in this system and thus no curling magnetization driven by it. This situation resembles the one at the termination of a wire where part of the surface charges are converted in volume charges of the same sign that extend to a certain axial distance instead of just at the interface, leading to a curling state and a decrease of the dipolar energy \cite{bib-ARR1979}. 
 
For the case of a finite lower magnetization chemical modulation, the repulsion between magnetic charges would extend the cloud of magnetic volume charges outwards of the chemical modulation. In Fig. \ref{fig:FIGS5} a) the schematics of charges can be observed. In this case axial magnetization would appear at the modulation and curling magnetization next to the interfaces. Axial magnetization at the modulation is allowed due to the apperance of volume charges of opposite sign that screen the surface charges. Fig. \ref{fig:FIGS5} b) shows micromagnetic simulations of the micromagnetic state at rest of a 90 nm diameter ${\text{Fe}_{80}\text{Ni}_{20}}$ nanowire with a 100 nm long ${\text{Fe}_{20}\text{Ni}_{80}}$ chemical modulation.  The color code represents the value of axial magnetization and the arrows the direction of magnetization. Axial magnetization at the ${\text{Fe}_{20}\text{Ni}_{80}}$ modulation and curling magnetization at the interfaces is confirmed.

\bibliography{Fruche8,Lauracomp}